\documentclass[showpacs,floatfix,preprint]{revtex4-1}
\usepackage{graphicx,psfrag,amsmath,amssymb,amsfonts,latexsym,color,epsf,graphpap,dcolumn}
\newcommand{\pp}{{\mathbf p}_\parallel}
\newcommand{\kp}{{\mathbf k}_\parallel}

\begin{document}
\title{On the Derivative Expansion for the Electromagnetic Casimir Free
Energy at High Temperatures}
\author{C.~D.~Fosco$^{a}$, F.~C.~Lombardo$^{b}$, F.~D.~Mazzitelli$^{a}$}
\affiliation{$^a$Centro At\'omico Bariloche and Instituto Balseiro, 
Comisi\'on Nacional de Energ\'\i a At\'omica, R8402AGP Bariloche, Argentina.\\
$^b$Departamento de F\'\i sica {\it Juan Jos\'e
Giambiagi}, FCEyN UBA and IFIBA CONICET-UBA, Facultad de Ciencias Exactas y Naturales,
Ciudad Universitaria, Pabell\' on I, 1428 Buenos Aires, Argentina}

\date{today}

\begin{abstract}
\noindent 
We study the contribution of the thermal zero modes to the Casimir free
energy, in the case of a fluctuating electromagnetic (EM) field in the presence
of real materials described by frequency-dependent, local and isotropic
permittivity ($\epsilon$) and permeability ($\mu$) functions. 

Those zero modes, present at any finite temperature, become dominant at
high temperatures, since the theory is dimensionally reduced.  Our work,
within the context of the Derivative Expansion (DE) approach, focusses on
the emergence of non analyticities in that dimensionally reduced theory.
We conclude that the DE is well defined whenever the function
$\Omega(\omega)$, defined by $[\Omega(\omega)]^2 \equiv
\omega^2\epsilon(\omega)$, vanishes in the zero-frequency limit, for at
least one of the two material media involved.
\end{abstract}
\pacs{12.20.Ds, 03.70.+k, 11.10.-z}
\maketitle
\section{Introduction}\label{sec:intro}
Casimir forces are one of the most remarkable macroscopic manifestations of
the vacuum fluctuations of the EM field~\cite{books}.
The precision achieved in recent experiments to measure those forces has
encouraged further work to obtain more detailed theoretical predictions for
them; in particular, it has become increasingly clear that the geometry and electromagnetic
properties of the materials are the two most important aspects that must be
taken into account for those predictions to be accurate. 

In spite of the intense activity in this field of research, comparison
between theory and experiment is not yet entirely satisfying, what leaves
some room for debate~\cite{Mostep2015}.  Among the most widely used tools
to tackle these issues, Lifshitz formula~\cite{Lifshitz} occupies a
prominent place.  Indeed, on the one hand it can be used as the starting
point for computing the Casimir force away from the idealized, perfect
conductivity case. On the other, it can also be applied to incorporate
non-trivial geometrical effects, at least when the surfaces involved are
close to each other and smooth. Indeed, Lifshitz formula, originally meant
just for flat and parallel slabs may also be thought of as the leading term, the so
called Proximity Force Approximation~\cite{Derjaguin} (PFA), in a
Derivative Expansion (DE) of the free energy, regarded as a functional of
the shape of the function(s) defining the surfaces involved~\cite{denos,
deothers}. 
In the language of Quantum Field Theory, the DE may be consistently viewed
as a low-momentum expansion of the vertex functions, with the PFA playing
the role of leading term, involving the summation of an infinite number of  vertex
functions at zero-momentum. The next to leading order term, in turn,
requires the knowledge of vertex functions at the second order in momentum.
A similar approach has also been applied in~\cite{Bimonte:2014mma} to find
an interesting approximation for the Casimir-Polder interaction force,
assuming that the surfaces are gently curved.

Regarding the EM properties of the media involved, it is usually sufficient
to use a version of Lifshitz formula which describes them by means of an
isotropic, frequency dependent permittivity $\epsilon(\omega)$ and, when
magnetic properties are relevant, also its permeability $\mu(\omega)$.
Note, however, that even though Lifshitz formula has been derived using a
variety of approaches~\cite{Lifshitzderiv}, its applicability to
dissipative systems is still debated \cite{Mostep2015,Mostep2009,Guerout2014}. 

It has been shown that, in some particular situations, non-analyticities in
the expansion of the vertex functions could produce non-local NTLO
corrections to the PFA within the context of the DE.  This has been
explicitly shown for quantum scalar fields at non zero temperature in the
presence of perfect mirrors described by Neumann boundary
conditions)~\cite{deTfin}. This also holds true for a real scalar field
with Neumann conditions in $2+1$ dimensions, albeit it can be shown that the
non-analyticity can be cured (in a concrete model) by introducing a small departure from perfect
Neumann conditions~\cite{Neumann2+1}.   

In this article, we analyse the emergence of non-analyticities in the DE
for real materials which, based on the insight from our previous
work~\cite{Neumann2+1}, should come from contributions due to the
dimensionally reduced massless modes, which appear as the zero frequency
terms in a Matsubara expansion of the fields.  To carry out such an
analysis, it is convenient to have a general formula for the DE,
corresponding to the free energy for the EM field in the presence of media
with realistic properties.  This kind of formula has been used (although
not explicitly displayed)  by other authors in a previous
work~\cite{deothers}. There, the Casimir free energy has been evaluated
using the scattering matrix approach, based on earlier results about the
$S$-matrix elements for the scattering of electromagnetic waves, computed
in a perturbative expansion in the departure from the planar surface
case~\cite{Voronovich}. 

It is our aim in this work to carry out an analysis of the validity of the
DE using a different approach, for a closely related physical system: we
incorporate the possibility for the media to have non-trivial magnetic
properties.  On the other hand, our approach to the construction of the DE
proceeds along different lines: we use a functional integral approach and
the Matsubara formalism to single out the zero mode contributions from the
very beginning. We then apply the DE approach, performing an independent
perturbative calculation in that dimensionally reduced
theory. We find conditions for the DE to be well-defined, and single out models
where those conditions are not met.  

In the first principles functional approach that we follow, the derivation
dwells on a subtle point related to the choice of the temporal gauge for
the EM field at a non-zero temperature~\cite{ZinnJustin}, which
allows for a clean isolation of the problematic zero mode, the origin of the
``plasma-Drude'' controversy in the calculation of the Casimir force at a non
vanishing temperature~\cite{Mostep2015, Mostep2009,MiltonT}. 
Remarkably, it is precisely this zero mode the same which, depending on the
case considered, may or not exhibit a non-analytic behavior as a function of the
external momenta of the generalized vertices.

The structure of this article is as follows: in Sect.~\ref{sec:thesys} we
present the system that we want to study and the main assumptions made in
order to calculate its Casimir free energy, having in mind its treatment
within the context of the DE approach.  
Then, in Sect.~\ref{sec:results} we present explicit results on the
zero-mode contributions to the free energy, for media described by
non trivial permittivities or permeabilities.  In Sect.~\ref{sec:analyt} we
study the eventual emergence of non-analyticities in the NTLO correction
of the PFA. 
Finally, in Sect.~\ref{sec:conc} we present our conclusions.
Some technical but nevertheless relevant details of the calculations are
presented in Appendix A.

\section{The system}\label{sec:thesys}
\subsection{Geometrical set-up, definitions and conventions} 
The geometry of the systems that we consider can be determined by
specifying just two surfaces, denoted by $L$ and $R$. Our construction will
begin with cases where they are the boundaries between two spatial regions,
filled by material media. 

The surfaces $L$ and $R$ are, respectively, defined by \mbox{$x_3 = 0$}
and \mbox{$x_3 = \psi(x_1,x_2)$}, where $\psi$ is a smooth function of
two Cartesian coordinates on an $x_3={\rm constant}$ plane, for which we
will adopt the shorthand notation \mbox{${\mathbf x}_\parallel \equiv
(x_1,x_2)$}.  

A real medium will be described by introducing isotropic, spatially local
permittivities and permeabilities (in other words, they are defined by
scalar functions, since the corresponding tensors are proportional to the identity
matrix). 

The perfect-conductor case will be obtained as a particular limit.

\subsection{Free energy and partition function}
Since we deal with a quantum field at a finite temperature, it is
convenient to introduce its free energy
$F(\psi)$,  a function of the inverse temperature $\beta =
T^{-1}$ (in our conventions, Boltzmann's constant $k_B \equiv 1$) and a
real functional, i.e., a real-valued function, of $\psi({\mathbf
x}_\parallel)$. For the sake of notational clarity, the dependence on
$\beta$ is not explicitly shown.  $F$  may be written in terms of the partition
function, ${\mathcal Z}(\psi)$, as follows:
\begin{equation}
	F(\psi)\,=\,-\frac{1}{\beta}\,\log\big[\frac{{\mathcal
	Z}(\psi)}{{\mathcal Z}_0}\big]\;,
\end{equation}
where the denominator, ${\mathcal Z}_0$, denotes the partition
function for the free EM field, i.e., in the absence of media. The effect
of that denominator is to subtract the free energy of a free Bose gas of
photons in the absence of the mirrors, which does not contribute to the
force between them.  
There are other contributions to the free energy which are independent of the
distance between the two media, namely, that are invariant under
$\psi({\mathbf x}_\parallel) \to \psi({\mathbf x}_\parallel) + b$ ($b
\equiv {\rm constant}$). They can, usually in a rather straightforward way,
be identified with self-energies of the mirrors. That is, contributions which
measure the energy of the distorted vacuum corresponding to the EM field in
the presence of just one medium (taking the zero-point energy of the EM
field as reference). Since we are ultimately interested in the calculation
of the part of the free energy which is responsible for the (normal) force
between the two media, those terms will be discarded. 

In the Matsubara (imaginary-time) formalism, a functional integral expression
for the partition function ${\mathcal Z}(\psi)$ can be constructed by
integrating over  field configurations depending on the spatial
coordinates ${\mathbf x}$ and the imaginary time $x_0 \equiv \tau$. The
fields are periodic, with period $\beta$, in the imaginary time. Denoting
by $A=(A_\mu)$, ($\mu=0,1,2,3$) the $4$-potential in Euclidean spacetime, the form of
${\mathcal Z}(\psi)$ is as follows:
\begin{equation}\label{eq:defzbeta}
{\mathcal Z}(\psi) \;=\; \int \big[{\mathcal D}A\big] \;
e^{-{\mathcal S}_{\rm inv}(A)}
\end{equation} 
where ${\mathcal S}_{\rm inv}(A)$ is the gauge-invariant action for $A$, with
gauge transformations given by \mbox{$A_\mu(x) \to A_\mu(x) +
\delta_\lambda A_\mu(x)$}, \mbox{$\delta_\lambda A_\mu(x) =
\partial_\mu\lambda(x)$}.  ${\mathcal D}A$ is the unconstrained
functional-integration measure, while $\big[{\mathcal D}A\big]$ is used to
denote that measure after gauge fixing.
 
In terms of the components of the field strength tensor
$F_{\mu\nu}=\partial_\mu A_\nu - \partial_\nu A_\mu$, the form of the
gauge-invariant action is:
\begin{align}\label{eq:defsinv}
{\mathcal S}_{\rm inv}(A)\;=\;\int_0^\beta & d\tau \int_0^\beta d\tau' \int d^3{\mathbf x} \,
\big[  \frac{1}{2} \, F_{0j}(\tau,{\mathbf x})
\epsilon(\tau-\tau', {\mathbf x}) F_{0j}(\tau',{\mathbf x}) \nonumber\\ 
 + & \frac{1}{4} F_{ij}(\tau,{\mathbf x})
\mu^{-1}(\tau-\tau', {\mathbf x})  F_{ij} (\tau,{\mathbf x}) \big] \;,
\end{align}
where indices from the middle of the Roman alphabet run over spatial
indices (Einstein summation convention has been adopted), and
$\epsilon(\tau-\tau' , {\mathbf x})$ and $\mu(\tau-\tau' , {\mathbf x})$ 
denote the Euclidean versions of the permittivity and permeability,
respectively ($\mu^{-1}$ is the inverse integral kernel of $\mu$, with
respect to its time-like arguments). 
Space locality of those response functions has been assumed implicitly.

On account of our assumptions on the geometry of the system, we have that:
\begin{eqnarray}
\epsilon(\tau-\tau', {\mathbf x}) &=& 
\theta(-x_3) \,\epsilon_L(\tau-\tau') \,+\,
\theta(x_3) \,\theta(\psi({\mathbf x}_\parallel) - x_3)+
\theta(x_3-\psi({\mathbf x}_\parallel)) 
\epsilon_R(\tau-\tau')\nonumber \\
\mu(\tau-\tau', {\mathbf x}) &=& 
\theta(-x_3) \,\mu_L(\tau-\tau') \,+\,
\theta(x_3) \,\theta(\psi({\mathbf x}_\parallel) - x_3)+
\theta(x_3-\psi({\mathbf x}_\parallel)) \,\mu_R(\tau-\tau'),
\end{eqnarray}
where $\epsilon_{L,R}(\tau-\tau')$ and $\mu_{L,R}(\tau-\tau')$ characterize
the permittivity and permeability of the respective mirror. 

Note that the free (vacuum) form of the action is
\begin{equation}\label{eq:defsinvac}
{\mathcal S}_{\rm inv}^{(vac)}(A)\;=\;\frac{1}{4}\,\int d^4 x \;
F_{\mu\nu} F_{\mu\nu}  \;,
\end{equation}
where $d^4x \equiv d\tau d{\mathbf x}$, the integral over $\tau$ goes from
$0$ to $\beta$ (the fields being periodic, any interval of length
$\beta$ can be used as the extent of the imaginary time coordinate to
define the action). 

\subsection{Matsubara modes and gauge fixing}
The action is invariant under translations in the imaginary time: $\tau \to
\tau \,+\, {\rm constant}$. This suggests the use of  mixed Fourier
transformations for the fields, as well as for the response functions:
\begin{eqnarray}\label{eq:fou1}
A_\mu (\tau,{\mathbf x}) \;=\; \frac{1}{\beta} \,
\sum_{n=-\infty}^{+\infty} \widetilde{A}_\mu^{(n)}({\mathbf x}) \, e^{i \omega_n
\tau} \nonumber\\
\epsilon(\tau-\tau',{\mathbf x}) \;=\; \frac{1}{\beta} \,
\sum_{n=-\infty}^{+\infty} \widetilde{\epsilon}^{(n)}({\mathbf x}) \, e^{i \omega_n
(\tau-\tau')} \nonumber\\
\mu(\tau-\tau',{\mathbf x}) \;=\; \frac{1}{\beta} \,
\sum_{n=-\infty}^{+\infty} \widetilde{\mu}^{(n)}({\mathbf x}) \, e^{i \omega_n
(\tau-\tau')} 
\end{eqnarray}   
where $\omega_n \equiv \frac{2\pi n}{\beta}$ ($n \in {\mathbb Z}$) are the
Matsubara frequencies. The gauge transformations can be represented in
Fourier space; to that end, note that, since the field $A_\mu(\tau,{\mathbf
x})$ is periodic, so must be $\delta_\lambda A_\mu(\tau,{\mathbf x})$. 
Then we can expand this object in terms of a Fourier series,
\begin{equation}
\delta_\lambda A_\mu(\tau,{\mathbf x})\;=\;
\frac{1}{\beta}\, \sum_{n=-\infty}^{+\infty} c_\mu^{(n)}({\mathbf x}) e^{i
\omega_n \tau} \;.
\end{equation}
Note, however, that $\lambda(\tau,{\mathbf x})$ needs not to be periodic.
It may be seen that its more general form is as follows:
\begin{equation}
\lambda(\tau,{\mathbf x}) \;=\; C \, \tau \, +\, D \, + \, \lambda_P(\tau,{\mathbf x}) 
\end{equation}
where $C$ and $D$ are constants, and $\lambda_P(\tau,{\mathbf x})$ is periodic:
\begin{equation}
\lambda_P(\tau,{\mathbf x})\;=\; \frac{1}{\beta}\,
\sum_{n=-\infty}^{+\infty} \widetilde{\lambda}_P^{(n)}({\mathbf x}) e^{i
\omega_n \tau} \;.
\end{equation}
Hence, the full set of gauge transformations for the gauge field may be
written, using a Fourier transform for the time arguments, as follows:
\begin{eqnarray}\label{eq:gaugefou}
\delta_\lambda \widetilde{A}_j^{(n)}({\mathbf x}) &=&  \partial_j
\widetilde{\lambda}_P^{(n)}({\mathbf x}) \nonumber\\
\delta_\lambda \widetilde{A}_0^{(n)}({\mathbf x}) &=&  
\left\{ 
\begin{array}{ccc}
i \omega_n \,\widetilde{\lambda}_P^{(n)}({\mathbf x}) & {\rm if} & n \neq 0 \\
C & {\rm if}  & n = 0 
\end{array}
\right.\;.
\end{eqnarray}
From the explicit form above  for the gauge transformations in the same
`mixed' Fourier representation, we see that one possible gauge fixing
condition, which we will adopt throughout this article, is the {\em
temporal\/} gauge, which in this finite-temperature setup is given by:
\begin{equation}\label{eq:tempgauge}
\widetilde{A}_0^{(n)}({\mathbf x}) \;=\; 0 \;\;,\;\;\;\; \forall n \neq 0 \;.
\end{equation}

We then write the thermal partition function using the mixed Fourier
representation. Using the properties: $\widetilde{A}_\mu^{(n)*} =
\widetilde{A}_\mu^{(-n)}$, $\widetilde{\epsilon}^{(-n)} =
\widetilde{\epsilon}^{(n)}$ and $\widetilde{\mu}^{(-n)} =
\widetilde{\mu}^{(n)}$, it is rather straightforward to see that it
may be written as an infinite product of (decoupled) integrals, involving
the zero mode, and just the positive (or, alternatively, negative) modes: 
\begin{equation}\label{eq:zprod}
{\mathcal Z}(\psi) \;=\; \prod_{n=0}^\infty {\mathcal Z}^{(n)}(\psi) \;
\end{equation}
where
\begin{equation}
{\mathcal Z}^{(0)}(\psi) \;=\; \int [{\mathcal
D}\widetilde{A}_0^{(0)}{\mathcal D}\widetilde{A}_j^{(0)}]\;e^{- {\mathcal
S}^{(0)}(\widetilde{A}_0^{(0)}, \widetilde{A}_j^{(0)})}
\end{equation}
and, for $n \geq 1$,
\begin{equation}
{\mathcal Z}^{(n)}(\psi) \;=\; \int [{\mathcal
D}\widetilde{A}_j^{(n)}{\mathcal D}\tilde{A}_j^{(n)*}]\;e^{- {\mathcal
S}^{(n)}(\widetilde{A}_j^{(n)}, \widetilde{A}_j^{(n)*}) }
\end{equation}
where ${\mathcal S}^{(0)}$ involves two real fields: one of them is the zero-frequency
component of $\widetilde{A}_0$, which behaves as a scalar in the Euclidean $2+1$
dimensional space. The other, corresponding to the zero-mode for
$\widetilde{A}_j$, is a real vector field:
\begin{equation}
{\mathcal S}^{(0)}(\widetilde{A}_0^{(0)}, \widetilde{A}_j^{(0)}) \,=\, 
\frac{1}{\beta}
\int d^3{\mathbf x} \,
\big[ \frac{1}{2} \, \widetilde{\epsilon}^{(0)}({\mathbf x}) 
(\partial_j \widetilde{A}_0^{(0)})^2 
\,+\, 
\frac{1}{4 \,\widetilde{\mu}^{(0)}({\mathbf x})} (\widetilde{F}_{jk}^{(0)})^2 
\,+\, \frac{1}{2} \, \Omega_0^2({\mathbf x}) (\widetilde{A}_j^{(0)})^2
\big]
\end{equation}
where we have introduced:
\begin{equation}
\Omega_0^2({\mathbf x}) \,\equiv\, \lim_{n\to 0} \,
\big[ \omega_n^2 \,\widetilde{\epsilon}^{(n)}({\mathbf x}) \big]\;.
\end{equation}
Note that $\Omega_0$ vanishes for a dielectric and also for a metal
described by the Drude model. On the other hand, it equals 
the plasma frequency for a metal described by the plasma model.  

Since, by assumption,  there is vacuum between the two mirrors,
$\widetilde{\epsilon}^{(0)} = \widetilde{\mu}^{(0)} =
1$, and \mbox{$\Omega_0^2 = 0$} in that region. 
Thus the scalar and vector fields behave there as a
free massless scalar and a free gauge field, respectively. 
In each one of the regions occupied by a mirror, the scalar field Lagrangian is multiplied by a
constant (in a wave-function renormalization fashion), while the vector
field also has a constant Proca-like mass (each constant is determined by
the properties of the medium on the mirror considered). Besides the latter
also has a factor similar to the scalar field one, but determined by the permeability.

As a consequence of the temporal gauge choice, there is no scalar field
mode in the ${\mathcal S}_{n>0}$ terms. They always involve just a complex vector field:
\begin{equation}
{\mathcal S}^{(n)}(\widetilde{A}_j^{(n)}, \widetilde{A}_j^{(n)*})\,=\,
\frac{1}{\beta}
\int d^3{\mathbf x} \,
\big( 
\frac{1}{2} |\widetilde{F}_{jk}^{(n)}|^2 
\,+\,  \Omega_n^2({\mathbf x}) |\widetilde{A}_j^{(0)}|^2
\big) \;,
\end{equation}
with a Proca-like mass \mbox{$\Omega_n^2({\mathbf x})
=\widetilde{\epsilon}^{(n)} ({\mathbf x}) \omega_n^2$} which, contrary to
what happened for the zero mode, is non-vanishing between the mirrors.
Note that, except for the $n = 0$ mode, each term corresponds to combining
two different Matsubara modes, namely, $n$ and $-n$ into the action for a
complex field. This is possible because of the reality of the permittivity
in Fourier space, which is in turn a reflection of the parity in its time
argument.

In any case, the free energy will be obtained as a sum of infinite terms,
each one corresponding to a given value of the index $n$,
\begin{equation}
F(\psi)\;=\; \sum_{n=0}^\infty \, F^{(n)}(\psi) \;,
\end{equation}
where 
\begin{equation}
F^{(n)}(\psi) \;=\;-\frac{1}{\beta}\log\big[ \frac{{\mathcal
Z}^{(n)}(\psi)}{{\mathcal Z}_0^{(n)}} \big]\;.
\end{equation}

In the special case of $n=0$, we may split the free energy according to its
origin being the scalar ($s$) or vector ($v$) fields:
\begin{equation}
	F^{(0)} \;=\; F_s(\psi) \,+\, F^{(0)}_v(\psi) \;,
\end{equation}
while for $n \geq 1$, we only have contributions originated in a vector
field $F^{(n)} \,=\,F^{(n)}_v$, $\forall n \geq 1$.
\section{Results for the zero mode free energies}\label{sec:results}
This contribution to the free energy is composed of two independent terms,
$F^{(0)}=F_s +F^{(0)}_v$, each one of them can be obtained from the
calculation of a path integral over an (unconstrained) real field, namely:
\begin{equation}
e^{- \beta \, F_s(\psi)} \,=\, \int {\mathcal D} \widetilde{A}_0^{(0)} \,
e^{- \frac{1}{2 \beta}
\int d^3{\mathbf x} \, \widetilde{\epsilon}^{(0)}({\mathbf x}) 
(\partial_j \widetilde{A}_0^{(0)})^2 }
\end{equation}
and 
\begin{equation}
e^{- \beta \, F^{(0)}_v(\psi)} \,=\, \int {\mathcal D}
\widetilde{A}_j^{(0)} \,
e^{- \frac{1}{\beta} \int d^3{\mathbf x} [ \frac{1}{4\,
\widetilde{\mu}^{(0)}({\mathbf x})} (\widetilde{F}_{jk}^{(0)})^2 
+ \frac{1}{2} \, \Omega_0^2({\mathbf x}) (\widetilde{A}_j^{(0)})^2]} \;.
\end{equation}
The functions $\widetilde{\epsilon}^{(0)}$, $\widetilde{\mu}^{(0)}$ and
$\Omega_0$ are model-dependent. Nevertheless, since by assumption we have a
vacuum between the plates, we can write:
\begin{eqnarray}
\widetilde{\epsilon}^{(0)}({\mathbf x}) &=& \epsilon_L \,
\theta(-x_3) \,+\,
\theta(x_3) \,\theta(\psi({\mathbf x}_\parallel) - x_3)+
\epsilon_R \,
\theta(x_3-\psi({\mathbf x}_\parallel)) \;,
\end{eqnarray}
\begin{eqnarray}
\widetilde{\mu}^{(0)}({\mathbf x}) &=& \mu_L \,
\theta(-x_3) \,+\,
\theta(x_3) \,\theta(\psi({\mathbf x}_\parallel) - x_3)+
\mu_R \,
\theta(x_3-\psi({\mathbf x}_\parallel)) \;,
\end{eqnarray}
and
\begin{eqnarray}
\Omega_0^2({\mathbf x}) &=& \Omega_L^2 \,
\theta(-x_3) \,+\, \Omega_R^2 \,
\theta(x_3-\psi({\mathbf x}_\parallel)) \;,
\end{eqnarray}
where we have introduced the constants:
\begin{eqnarray}
\epsilon_{L,R} &=& \lim_{n\to 0} \widetilde{\epsilon}_{L,R}^{(n)}
\nonumber\\ 
\Omega_{L,R}^2 &=& \lim_{n\to 0} [\omega_n^2 \widetilde{\epsilon}_{L,R}^{(n)}]
\;,
\end{eqnarray}
which are entirely determined by the parameters of the model used for the permittivity.

We take advantage of the fact that the two contributions above are
decoupled, to consider and present the corresponding results separately
below.
Before doing this, it is instructive to discuss the limit of perfectly 
conducting materials for the scalar and vector contributions. 
The $\widetilde{A}^{(0)}_0$ field behaves as a scalar, and when the
infinite permittivity limit is taken, its gradient inside the region
occupied by the mirror vanishes. Thus the field is constant in that region;
assuming that the conductors are grounded, that constant vanishes, so that the field
itself is zero. 
That is, this field becomes a Dirichlet mode, corresponding to the
transverse magnetic (TM) EM mode. 

The vector zero mode, on the other hand, behaves as an EM field in $2+1$
dimensions. If $\Omega_L$, say, tends to infinity, then the that  EM field
will vanish identically on the region occupied by that mirror. It then has
perfect conductor boundary conditions at $x_3 = 0$. (Of course, the same
will happen on the other surface if the corresponding constant tends to infinity.)
But we have shown this to be equivalent to a real scalar field with
Neumann conditions~\cite{Neumann2+1}. Therefore, this is a Neumann
mode, corresponding to the transverse electric (TE) EM mode. 

\subsection{The scalar zero-mode contribution $F_s(\psi)$}
We first note that a formal result of the functional integral for this mode
can be written in terms of the determinant of an operator ${\mathcal K}_s$:
\begin{eqnarray}\label{eq:defgs}
e^{- \beta \, F_s(\psi)} &=& [\det {\mathcal K}_s ]^{-\frac{1}{2}} \nonumber\\
{\mathcal K}_s &=& - \partial_j \widetilde{\epsilon}^{(0)}({\mathbf x}) \partial_j \;,
\end{eqnarray}
where we have neglected a global constant which does not contribute to the
interaction energy between the mirrors.

Within the DE approach to the second order, $F_s$ is given by an expression
with the form:
\begin{equation}\label{des}
[F_s(\psi)]_{DE} \;=\; \int d^2{\mathbf x}_\parallel \,
\Big[ V_s(\psi) \,+\, Z_s(\psi) \partial_a \psi({\mathbf x}) \partial_a
\psi({\mathbf x})\Big] \;.
\end{equation}
The functions $V_s$ and $Z_s$ may be determined from the first and third terms
in the expansion of $F_s(a + \eta)$ in powers of $\eta$: $F_s(a + \eta)=
F_{s,0}(a) + F_{s,1}(a,\eta) + F_{s,2}(a,\eta)+ \ldots $, as described in our previous works
\cite{denos}. 
\subsubsection{Zeroth order term}
The first term, which contains no derivatives, allows we to construct the
corresponding `potential' in the DE,
\begin{equation}
V_s(a) \,=\, \frac{1}{L^2} \, F_{s,0}(a)  \;,
\end{equation}
where $L^2$ denotes the area of each surface ($L \to \infty$).
From (\ref{eq:defgs}), this implies
\begin{equation}
V_s(a) \,=\, \frac{1}{2 \beta} \,\int \frac{d^2{\mathbf
k}_\parallel}{(2\pi)^2} \, \log \det \big[{\mathcal K}_s|_{\psi({\mathbf
x}_\parallel = a}\big]\;.
\end{equation}

This contribution can be evaluated exactly, taking advantage of the fact
that it corresponds to a system where $\widetilde{\epsilon}^{(0)}({\mathbf
x})$ depends only on $x_3$, by
using, for example, the Gelfand-Yaglom theorem approach \cite{Ccappa}. Since the calculation is already a standard
one, we omit the detalis.  The result is:
\begin{equation}
\label{Vscal}
V_s(a) \,=\, \frac{1}{2 \beta} \,\int \frac{d^2{\mathbf
k}_\parallel}{(2\pi)^2} \, \log \big( 1 - r_L r_R e^{- 2 |{\mathbf
k}_\parallel| a}\big)\;,
\end{equation}
where 
\begin{equation}\label{rdir}
r_{L,R} \equiv \frac{1 - \epsilon_{L,R}}{1 +
\epsilon_{L,R}}\, . 
\end{equation}
As expected, in
the limit  $\epsilon_{L,R}\to\infty$ one obtains the result corresponding to a scalar field in $2+1$ dimensions
satisfying Dirichlet boundary conditions times $1/\beta$ \cite{deTfin}. 

\subsubsection{First order term}
The first order term, $F_{s,1}$, is not required in order to determine the
DE; however, we calculate it in order to have a consistency check for some
of the ingredients we use in our work. 
We first note that
$F_s(\psi)$ can be conveniently written in an equivalent form where the
zeroth order term is extracted explicitly,
\begin{equation}
e^{-\beta F_s} \,=\, e^{-\beta F_{s,0}} \;  e^{-\beta F_{s,I}}
\end{equation}
where $F_{s,I}$ contains all the perturbative corrections:
\begin{equation}
F_{s,I} \;=\; - \frac{1}{\beta} \, 
\log \big\langle e^{- {\mathcal S}_{s,I}}\big\rangle \;.
\end{equation}
The average symbol is given by a functional integral with the
action corresponding to $\eta \equiv 0$:
\begin{equation}
\big\langle \ldots \big\rangle \;=\; 
\frac{\int {\mathcal D} \widetilde{A}_0^{(0)} \ldots e^{- {\mathcal S}_{s,0}(\widetilde{A}_0^{(0)})}}{\int {\mathcal D} \widetilde{A}_0^{(0)} e^{- {\mathcal S}_{s,0}(\widetilde{A}_0^{(0)})}} \;. 
\end{equation}
Here,
\begin{equation}
{\mathcal S}_s \;=\; {\mathcal S}_{s,0}\,+\, {\mathcal S}_{s,I} \;
\end{equation}
and
\begin{equation}
{\mathcal S}_{s,0} \;\equiv \;  {\mathcal S}_s \Big|_{\psi({\mathbf
x}_\parallel)=a} \;.
\end{equation}
Then, the first order term becomes:
\begin{equation}
F_{s,1} \;=\;\frac{1}{\beta} \,\big\langle {\mathcal S}_{s,1}\big\rangle \;,
\end{equation}
where the first-order term in the action, ${\mathcal S}_{s,1}$, is given by
\begin{equation}
{\mathcal S}_{s,1} \;=\; \frac{1-\epsilon_R}{2 \beta} \int d^3{\mathbf x}
\, \delta(x_3 - a) \, |\nabla \widetilde{A}_0^{(0)}({\mathbf x})|^2 \;.
\end{equation}
Thus, 
\begin{equation}
F_{s,1} \;=\;\frac{1 - \epsilon_R}{ 2\beta^2} \,\int d^3{\mathbf x}
\, \delta(x_3 - a) \,\Delta_{jj}({\mathbf x}_\parallel, {\mathbf
x}_\parallel)  
\end{equation}
where we have introduced the objects:
\begin{equation}
\Delta_{jl}({\mathbf x}_\parallel , {\mathbf y}_\parallel) \;\equiv\;
\Delta_{jl}({\mathbf x}_\parallel - {\mathbf y}_\parallel) \;=\;
\langle \partial_j \widetilde{A}_0^{(0)}({\mathbf x})
\partial_l\widetilde{A}_0^{(0)}({\mathbf
y})\rangle\big|_{x_3, y_3 \to a}\;,
\end{equation}
On the other hand, the correlation function which coincidence limit appears
above, is ill-defined when one of the indices $j, l$ equals $3$. Indeed,
the derivatives of the field have a discontinuity whenever the permittivity
has a jump. We then introduce in ${\mathcal S}_{s,1}$, the point split
action, before evaluating the average:
\begin{equation}
{\mathcal S}^\eta_{s,1} \;=\; \frac{1-\epsilon_R}{2 \beta} \int d^3{\mathbf x}
\, \partial_j\widetilde{A}_0^{(0)}({\mathbf x}_\parallel, a+\eta) 
\partial_j\widetilde{A}_0^{(0)}({\mathbf x}_\parallel, a-\eta)\;.
\end{equation}
and obtain the result for the free energy by taking the $\eta \to 0$ limit
after evaluating the average.

Thus, we see that the first-order term is given by:
\begin{equation}
F_{s,1} \;=\;\frac{1 - \epsilon_R}{ 2\beta^2} \,\int d^2{\mathbf
x}_\parallel \, \big[ \Delta^{+-}_{aa}({\mathbf x}_\parallel, {\mathbf
x}_\parallel)  + \Delta^{+-}_{33}({\mathbf x}_\parallel, {\mathbf
x}_\parallel) \big] \, \eta({\mathbf x}_\parallel)\;,
\end{equation}
where
\begin{equation}
\Delta_{jl}^{+-}({\mathbf x}_\parallel - {\mathbf y}_\parallel) \;\equiv\;
\big[\langle \partial_j \widetilde{A}_0^{(0)}({\mathbf x})
\partial_l\widetilde{A}_0^{(0)}({\mathbf
y})\rangle\big|_{x_3 \to a^+, y_3 \to a^-}\;\equiv\; \int \frac{d^2{\mathbf
k}_\parallel}{(2\pi)^2} \, \widetilde{\Delta}^{+-}_{jl}({\mathbf k}_\parallel)
\, e^{i {\mathbf k}_\parallel \cdot ({\mathbf x}_\parallel - {\mathbf
y}_\parallel)} \;.
\end{equation}
We note that the $\pm$ in $\Delta_{ab}$ ($a\neq3$, $b\neq3$) may be
omitted, since that object is well defined, continuous at $x_3=y_3=a$.

A rather straightforward calculation shows that, when both indices are different from $3$:
\begin{equation}
\widetilde{\Delta}_{ab}({\mathbf k}_\parallel) \;=\; \beta \,
\widetilde{\Delta}({\mathbf k}_\parallel) \, k_a k_b \;,
\end{equation}
with:
\begin{equation}\label{eq:defdelta}
\widetilde{\Delta}({\mathbf k}_\parallel)  = \frac{1}{2 |{\mathbf
k}_\parallel|} \, 
\Big\{ \frac{1}{\epsilon_R} \, + \, 
\frac{1}{ 1 - r_L r_R e^{- 2 |{\mathbf k}_\parallel| a}}  \big[
r_L (\frac{2}{ 1 + \epsilon_R} + \frac{r_R}{\epsilon_R} ) e^{- 2 |{\mathbf k}_\parallel| a}
- \frac{r_R}{\epsilon_R} \big]\Big\} \;.
\end{equation}
On the other hand,
\begin{equation}\label{eq:defdel33}
\widetilde{\Delta}^{+-}_{33}({\mathbf k}_\parallel) \;=\;
\widetilde{\Delta}^{-+}_{33}({\mathbf k}_\parallel) \;=\; - \beta \,
\epsilon_R \, |{\mathbf k}_\parallel| \, \Big[ \frac{1}{\epsilon_R} \, 
-\, \widetilde{\Delta}({\mathbf k}_\parallel) |{\mathbf k}_\parallel| \Big]
\;,
\end{equation}
with the same $\widetilde{\Delta}$ as in the previous equation.

Thus, at the first order:
\begin{equation}
F_{s,1} \;=\;\frac{1 - \epsilon_R}{ 2\beta^2} \,\int d^2{\mathbf x}_\parallel  
\eta({\mathbf x}_\parallel) \, \int \frac{d^2{\mathbf k}_\parallel}{(2\pi)^2} \; 
\big[ \widetilde{\Delta}_{aa}({\mathbf k}_\parallel)
+\widetilde{\Delta}^{+-}_{33}({\mathbf k}_\parallel) \big] \;.
\end{equation}
Discarding $a$-independent terms, we find that the free energy per unit
area is:
\begin{equation}
\frac{F_{s,1}}{L^2} \;=\; \eta_0 \, \frac{r_L  r_R}{\beta} \, 
\int \frac{d^2{\mathbf k}_\parallel}{(2\pi)^2} \,
\frac{|{\mathbf k}_\parallel|}{e^{ 2 |{\mathbf k}_\parallel| a} - r_L r_R}
\;,
\end{equation}
with $\eta_0 \equiv \frac{\int d^2{\mathbf x}_\parallel  \eta({\mathbf
x}_\parallel)}{L^2}$.

The consistency check is completed by noting that this term agrees with $\eta_0
\frac{\partial}{\partial a} V_s(a)$.

\subsubsection{Second order term}
The second order term is obtained by collecting the second-order terms in
the expansion in powers of $\eta$.
Discarding self-energy like contributions, and with the same notations
introduced above, we see that it is, in principle, given by:
\begin{equation}\label{eq:fs2}
F_{s,2} \;=\; - \frac{1}{4 \beta^3} \, (1 - \epsilon_R)^2 \, 
\int d^2{\mathbf x}_\parallel 
\int d^2{\mathbf y}_\parallel 
\big[\langle \partial_j \widetilde{A}_0^{(0)}({\mathbf x}) \partial_k\widetilde{A}_0^{(0)}({\mathbf
y})\rangle\big|_{x_3 = y_3 =a}\big]^2 \, \eta({\mathbf x}_\parallel)
\eta({\mathbf y}_\parallel) \;.  
\end{equation}
The expression is, however, ill-defined.
Using the point splitting procedure, in the same form as in the first-order
calculation, to make sense of the ill-defined vertices, we obtain:
\begin{align}\label{eq:fs21}
F_{s,2} \;=\; & - \frac{1}{8 \beta^3} \, (1 - \epsilon_R)^2 \, 
\int d^2{\mathbf x}_\parallel \int d^2{\mathbf y}_\parallel 
\Big[ 
 2 \Delta_{ab}({\mathbf x}_\parallel - {\mathbf y}_\parallel)
\Delta_{ab}({\mathbf x}_\parallel - {\mathbf y}_\parallel) \nonumber \\
& +
\Delta^{++}_{33}({\mathbf x}_\parallel - {\mathbf y}_\parallel)
\Delta^{--}_{33}({\mathbf x}_\parallel - {\mathbf y}_\parallel) 
+
\Delta^{+-}_{33}({\mathbf x}_\parallel - {\mathbf y}_\parallel)
\Delta^{-+}_{33}({\mathbf x}_\parallel - {\mathbf y}_\parallel) \nonumber \\
&+ \Delta^{+-}_{3a}({\mathbf x}_\parallel - {\mathbf y}_\parallel)
\Delta^{-+}_{3a}({\mathbf x}_\parallel - {\mathbf y}_\parallel) 
+
\Delta^{++}_{3a}({\mathbf x}_\parallel - {\mathbf y}_\parallel)
\Delta^{--}_{3a}({\mathbf x}_\parallel - {\mathbf y}_\parallel) \Big] 
\eta({\mathbf x}_\parallel)  \eta({\mathbf y}_\parallel) \;.
\end{align}

This term, being quadratic in $\eta$, can be represented in (parallel)
Fourier space in terms of a kernel $f_s^{(2)}$: 
\begin{equation}
	F_s^{(2)}\,=\, \frac{1}{2} \int \frac{d^2k_\parallel}{(2\pi)^2}
	\,f_s^{(2)}(k_\parallel, a) \, |\tilde{\eta}(k_\parallel)|^2 \;,
\end{equation}
where $k_\parallel=\vert\kp\vert$
The function  $Z_s(\psi)$ in Eq.(\ref{des}) can be obtained from 
\begin{equation}\label{Zs}
Z_s(\psi)=\frac{1}{2}\frac{\partial f_s^{(2)}}{\partial k_\parallel^2}(0, \psi)\, .
\end{equation}
Note that, as $\epsilon_{R,L}$ are dimensionless, dimensional analysis implies that
$Z_s$ is $1/(\beta\psi^4)$ times a function of $\epsilon_{R,L}$.

The explicit form for $f_s^{(2)}$ becomes: 
\begin{align}
& f_s^{(2)}(k_\parallel, a) \;=\;  - \frac{1}{2\beta} \, (1 - \epsilon_R)^2 \, 
\int \frac{d^2{\mathbf p}_\parallel}{(2\pi)^2}  \, \Big\{
\widetilde{\Delta}({\mathbf p}_\parallel) \,\widetilde{\Delta}({\mathbf
p}_\parallel + {\mathbf k}_\parallel) \big[  ( {\mathbf p}_\parallel \cdot
({\mathbf p}_\parallel + {\mathbf k}_\parallel) )^2 \nonumber\\
 & - 2 \epsilon_R \,   |{\mathbf p}_\parallel | |{\mathbf
p}_\parallel + {\mathbf k}_\parallel|  \,  {\mathbf p}_\parallel \cdot
({\mathbf p}_\parallel + {\mathbf k}_\parallel)\big] \nonumber\\
& + \epsilon_R^2 \big( - \frac{1}{\epsilon_R} \, |{\mathbf p}_\parallel|  \, 
+\, \widetilde{\Delta}({\mathbf p}_\parallel) |{\mathbf p}_\parallel|^2 \big) 
\big( - \frac{1}{\epsilon_R} \, |{\mathbf p}_\parallel + {\mathbf
k}_\parallel|\, +\, \widetilde{\Delta}({\mathbf p}_\parallel+ {\mathbf
k}_\parallel) |{\mathbf p}_\parallel+ {\mathbf k}_\parallel|^2 \big] 
\Big\} \;.
\end{align}
Subtracting from $f_s^{(2)}$ its value at $a\to\infty$ we obtain,  in the particular case $\epsilon_R=\epsilon_L=\epsilon$
\begin{eqnarray}
\label{f2scal}
f_s^{(2)}(k_\parallel, a) & = &  - \frac{1}{2\beta} \, (1 - \epsilon)^2 \, 
\int \frac{d^2{\mathbf p}_\parallel}{(2\pi)^2}  \, \Big\{\frac{1}{ |{\mathbf p}_\parallel | |{\mathbf
p}_\parallel + {\mathbf k}_\parallel|}
\big(A_-({\mathbf p}_\parallel)A_-({\mathbf p}_\parallel+{\mathbf k}_\parallel)-\frac{1}{(1+\epsilon)^2}\big)\nonumber\\
&\times&  \big[  ( {\mathbf p}_\parallel \cdot
({\mathbf p}_\parallel + {\mathbf k}_\parallel) )^2 
  - 2 \epsilon \,   |{\mathbf p}_\parallel | |{\mathbf
p}_\parallel + {\mathbf k}_\parallel|  \,  {\mathbf p}_\parallel \cdot
({\mathbf p}_\parallel + {\mathbf k}_\parallel)\big] \nonumber\\
&+&  |{\mathbf p}_\parallel | |{\mathbf
p}_\parallel + {\mathbf k}_\parallel| \big(\epsilon^2 A_+({\mathbf p}_\parallel)A_+({\mathbf p}_\parallel+{\mathbf k}_\parallel)-\frac{1}{(1+\epsilon)^2}\big)\Big\}\, ,
\end{eqnarray}
where
\begin{equation}
A_\pm({\mathbf p}_\parallel)=\frac{1}{2\epsilon}\left(1\pm \frac{r(1-e^{-2\vert {\mathbf p}_\parallel\vert a})}{1-r^2 e^{-2 \vert{\mathbf p}_\parallel\vert a}}\right)\, .
\end{equation}
In the 
 limit $\epsilon\to\infty$, and omitting a term independent of ${\mathbf k}_\parallel$,  it reduces to
 \begin{equation}
 f_s^{(2)}(k_\parallel, a)  =   - \frac{2}{\beta} \, 
\int \frac{d^2{\mathbf p}_\parallel}{(2\pi)^2} \frac{ |{\mathbf p}_\parallel | |{\mathbf
p}_\parallel + {\mathbf k}_\parallel| }{(1-e^{-2\vert {\mathbf p}_\parallel\vert a})(e^{2\vert {\mathbf p}_\parallel+{\mathbf k}_\parallel\vert a}-1)}\, .
  \end{equation}
As expected, this result coincides with $1/\beta$ times the one obtained for a Dirichlet scalar field
in $2+1$ dimensions \cite{deTfin}.

It is interesting to remark that Eq.(\ref{f2scal}) can be obtained with a rather different approach based on the scattering formula
for the Casimir free energy \cite{deothers}, using standard perturbation theory for the analysis of the incidence of classical electromagnetic waves 
on rough surfaces 
of small slope \cite{Voronovich} (see Appendix A). The derivation presented here sheds light on the gauge fixing procedure in the functional integral:
the temporal gauge does not fix the zero-frequency component of $\widetilde A_0$, giving rise to the scalar contribution to the Casimir free energy. 

\subsection{The vector zero-mode contribution $F_v(\psi)$}
Again, within the DE approach to the second order, we will have for $F_s$
an expression with the form:
\begin{equation}
[F_v(\psi)]_{DE} \;=\; \int d^2{\mathbf x}_\parallel \,
\Big[ V_v(\psi) \,+\, Z_v(\psi) \partial_a \psi({\mathbf x}) \partial_a
\psi({\mathbf x})\Big] \;.
\end{equation}
The functions $V_v$ and $Z_v$ may be determined from the first and third terms
in the expansion of $F_v(a + \eta)$ in powers of $\eta$: $F_v(a + \eta)=
F_{v,0}(a) + F_{v,1}(a,\eta) + F_{v,2}(a,\eta)+ \ldots $.

In what follows we will consider separately  the case of a magnetic materials with a permittivity such that
$\Omega_0({\mathbf x})$ vanishes identically (for example, if the permittivity
has a regular zero-frequency limit) and the case of a non-magnetic materials with $\widetilde{\mu}^{(0)}=1$ and  
$\Omega_0({\mathbf x})\neq 0$. 

\subsubsection {The case $\widetilde{\mu}^{(0)}\neq 1$ and $\Omega_0=0$.}\label{magnetic}

In this situation, the Casimir free energy is given by:
\begin{equation}
e^{- \beta \, F^{(0)}_v(\psi)} \,=\, \int {\mathcal D}
\widetilde{A}_j^{(0)} \,
e^{- \frac{1}{4 \beta} \int d^3{\mathbf x}
\frac{1}{\widetilde{\mu}^{(0)}({\mathbf x})} \,
(\widetilde{F}_{jk}^{(0)})^2 }\;.
\end{equation}
In order to evaluate the functional integral, we apply a result
we obtained in a previous paper \cite{Neumann2+1}, whereby we have shown that the
calculation of the effective action for a gauge field in the presence of
imperfect mirrors could be mapped to a scalar field model. Indeed,
using the duality
\begin{equation}
\partial_j\phi \leftrightarrow\epsilon_{jik}\partial_i A_k\, ,
\end{equation}
the free energy can be written in terms of the scalar field as
\begin{equation}
e^{- \beta \, F^{(0)}_v(\psi)} \,=\, \int {\mathcal D} \phi
e^{- \frac{1}{ 2\beta} \int d^3{\mathbf x}
\frac{1}{\widetilde{\mu}^{(0)}({\mathbf x})} \, (\partial_j \phi)^2 }\;.
\end{equation}
Therefore, the result for the previous integral can be borrowed from the ones
we have already obtained for the scalar zero mode, by extending the results to $0<\epsilon<1$ 
and replacing
$\epsilon_{L,R} \to 1/\mu_{L,R}$. 

One can readily check that in the limit  $\mu_{L,R}\to \infty$ one recovers the TE mode of the perfect conductor case.
Indeed, taking the corresponding limit $\epsilon_{L,R} \to 0$ in Eq.(\ref{Vscal}) and (\ref{f2scal}) one obtains
 \begin{equation}
 \label{limitN}
 f_v^{(2)}(k_\parallel, a)  =   - \frac{2}{\beta} \, 
\int \frac{d^2{\mathbf p}_\parallel}{(2\pi)^2} \frac { ( {\mathbf p}_\parallel \cdot
({\mathbf p}_\parallel + {\mathbf k}_\parallel) )^2}{ |{\mathbf p}_\parallel | |{\mathbf
p}_\parallel + {\mathbf k}_\parallel| }\frac{1}{(1-e^{-2\vert {\mathbf p}_\parallel\vert a})(e^{2\vert {\mathbf p}_\parallel+{\mathbf k}_\parallel\vert a}-1)}\, ,
  \end{equation}
which is $1/\beta$ times the result for a Neumann scalar field in $2+1$ dimensions.

\subsubsection {The case $\widetilde{\mu}^{(0)}= 1$ and $\Omega_0\neq 0$.}

The `potential' term in the DE can again be evaluated exactly, since it
corresponds to a system where $\widetilde{\epsilon}^{(0)}({\mathbf x})$ 
depends only on $x_3$. An
application of the Gelfand-Yaglom theorem approach yields:
\begin{equation}
V_v(a) \,=\, \frac{1}{2 \beta} \,\int \frac{d^2{\mathbf
k}_\parallel}{(2\pi)^2} \, \log \big( 1 - \rho_L \rho_R e^{- 2 |{\mathbf
k}_\parallel| a}\big)\;,
\end{equation}
where we have introduced:
\begin{equation}
\label{rho}
 \rho_{L,R}({\mathbf k}_\parallel) = \frac{\Omega_{L,R}^2}{\left(|{\mathbf k}_\parallel|
+ \sqrt{|{\mathbf k}_\parallel|^2 + \Omega_{L,R}^2 }\right)^2}
\;=\;
\frac{\sqrt{|{\mathbf k}_\parallel|^2 + \Omega_{L,R}^2} - |{\mathbf k}_\parallel|}{\sqrt{|{\mathbf
k}_\parallel|^2 + \Omega_{L,R}^2} +|{\mathbf
k}_\parallel|} \;.
\end{equation}

Having already illustrated the computation of the Casimir free energy using the functional approach, 
and in order to avoid a rather lengthy calculation, 
for the second order term we will use the known results based on the scattering matrix approach.  As shown in Appendix A, 
the result for the second order kernel is, in the particular case $\Omega_L=\Omega_R=\Omega$: 
\begin{eqnarray}
\label{f2vv}
f_v^{(2)}(k_\parallel) &=& -\frac{2}{\beta} \, 
\int \frac{d^2{\mathbf p}_\parallel}{(2\pi)^2} \, 
\frac{\vert\pp\vert\rho^2({\mathbf p}_\parallel)}{g({\mathbf p}_\parallel)}e^{-2\vert{\mathbf p}_\parallel\vert a }
\Big\{\sqrt{\Omega^2+\vert{\mathbf p}_\parallel\vert^2}- \sqrt{\Omega^2+\vert{\mathbf p}_\parallel+{\mathbf k}_\parallel\vert^2}
\nonumber\\
&+&  \frac { ( {\mathbf p}_\parallel \cdot
({\mathbf p}_\parallel + {\mathbf k}_\parallel) )^2}{ |{\mathbf p}_\parallel |^2 |{\mathbf
p}_\parallel + {\mathbf k}_\parallel| g({\mathbf p}_\parallel+{\mathbf k}_\parallel)}
\Big\}\, ,
\end{eqnarray}
where we introduced the notation
\begin{equation}
\label{defg}
g({\mathbf k}_\parallel)=1-\rho^2({\mathbf k}_\parallel)e^{-2\vert{\mathbf k}_\parallel\vert a}\, .
\end{equation}
The perfect conductor limit is obtained as $\Omega\to\infty$.  In this limit, the difference of square roots in Eq.(\ref{f2vv}) vanishes, 
the reflection coefficient $\rho\to 1$, and the final result corresponds to a Neumann 
scalar field in $2+1$ dimensions (see Eq.({\ref{limitN}})).

\section{Analyticity of the Derivative Expansion}\label{sec:analyt}

In this section we analyze the structure of the NTLO correction to the
PFA. As repeatedly emphasized in  our previous works
\cite{denos,deTfin,Neumann2+1}, the DE is based on the low momentum
expansion of the kernels that appear in the perturbative evaluation of the
Casimir free energy,  when the shape of the interface between different
media is $\psi({\mathbf x}_\parallel)=a+\eta({\mathbf x}_\parallel)$ with
$a\ll \eta$ (in the present work, these kernels are $f_s^{(2)}$ and
$f_v^{(2)}$).  

If the kernels admit an expansion in powers of $k_\parallel^2$, the NTLO
correction to the PFA can be written locally in terms of derivatives of
$\psi$. In the opposite  case, when the low momentum behavior is
non-analytic in $k_\parallel^2$, the NTLO correction is  non-local. For
perfect conductors, we have shown that at finite temperature the TE
contribution contains a non-analytic contribution \cite{deTfin}. The origin
of the non-analyticity is in the zero-frequency mode, which behaves as a
scalar field with Neumann boundary conditions. Imperfect boundary
conditions may restore the analyticity of the kernels, at least in the toy
model considered in Ref.~\cite{Neumann2+1}.  A question that naturally
suggests itself is whether the departure from perfect conditions will
always restore the analyticity. To that end, the question we address
here is whether the boundary conditions for the EM field in
the presence of real materials at finite temperature make the kernels
analytic or not. In other words, if the resulting NTLO correction to PFA, for
realistic conditions at finite temperature, is spatially local or not. 

In order to analyze the low momentum behavior of $f_s^{(2)}$ and
$f_v^{(2)}$, we can proceed as follows: we first expand the integrand
defining both kernels (see Eqs.(\ref{f2scal}) and (\ref{f2vv})) in powers
of $k_\parallel^2$, and then look for eventual infrared divergences in the
term proportional to the first power of $k_\parallel^2$, which, when it is
well defined, constitutes the NTLO correction to PFA in the DE approach.

The expansion of $\vert{\mathbf p}_\parallel+{\mathbf k}_\parallel\vert$ in powers of $k_\parallel$ generates inverse powers of $p_\parallel$:
assuming that ${\mathbf k}_\parallel$ points in the $x$-direction we have
\begin{equation}
\vert{\mathbf p}_\parallel+{\mathbf k}_\parallel\vert = \sqrt{\vert\pp\vert^2+\vert\kp\vert^2+2\vert\pp\vert\kp\vert \cos\theta}\simeq
\vert\pp\vert+\cos\theta\vert\kp\vert+\frac{1}{2\vert\pp\vert}\sin^2\theta\vert\kp\vert^2+...\, ,
\end{equation}
so
the term proportional to $k_\parallel^2$ is inversely proportional to $p_\parallel$. This is one source of potential
infrared divergences. Additional inverse powers of $p_\parallel$ may appear when the reflection coefficients tend to one (see Eq.(\ref{defg})).

For the scalar kernel  (Eq.(\ref{f2scal})), the functions $A_\pm$ and their
derivatives contain at most one inverse power of $p_\parallel$. Therefore,
by power counting one can check that for any value  $0<\epsilon<\infty$
there are no infrared divergences and the NTLO correction to PFA is local
(the consideration of values $0<\epsilon<1$ is useful, since to analyze the case of
magnetic materials, we apply the duality $\mu\equiv
1/\epsilon$ ). In the limit $\epsilon\to \infty$ one obtains the Dirichlet
kernel, that does not contain infrared divergences.  In the opposite limit
$\epsilon\to 0$ the result corresponds to the Neumann kernel, which has an
infrared divergence when expanded in powers of $k_\parallel$. Therefore, we
conclude that the scalar contribution generates a well defined DE, and that
magnetic materials regulate the non-analyticity of the TE mode when
considering  very large (but not infinite) values of $\mu$.

\begin{figure}
\centering
\includegraphics[width=10cm , angle=0]{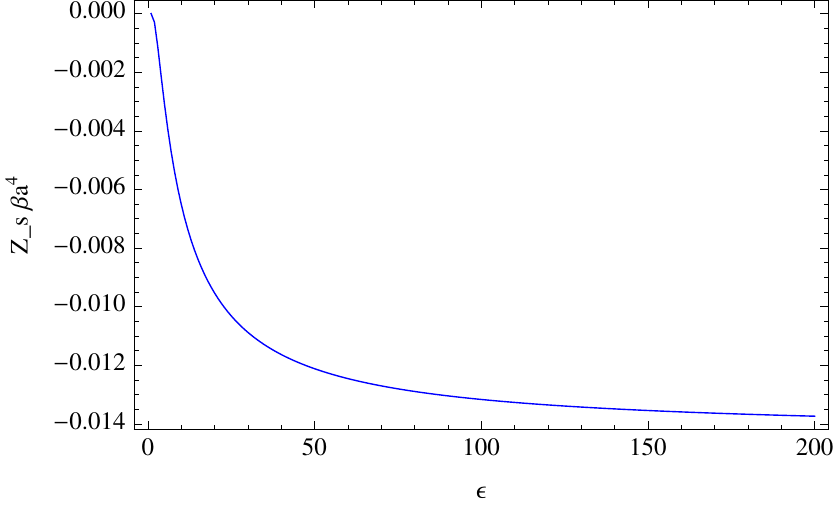}
\caption{Numerical evaluation of $Z_s(a)$ (in units of $\beta a^4$) given in  Eq.(\ref{Zs}) as a function of $\epsilon$. $Z_s$ vanishes when $\epsilon \rightarrow 1$ and tends to 
the value for a  2 + 1 Dirichlet mode (-0.0136) when $\epsilon\to\infty$. } \label{fig1}
\end{figure}

We have confirmed these results with numerical evaluations. In Fig.\ref{fig1} we plot
 $Z_s$, defined in  Eq. (\ref{Zs}), as a function of $\epsilon$. $Z_s= 0 $ for $\epsilon = 1$ and tends to the 2+1 Dirichlet value \cite{deTfin}
 \begin{equation}
 Z_s^{D}=-\frac{\Gamma(3/2)[1+6\zeta(3)]}{12(4\pi)^{3/2}}\simeq -0.0136
 \end{equation}
for large values of $\epsilon$.  
 
\begin{figure}
\centering
\includegraphics[width=10cm , angle=0]{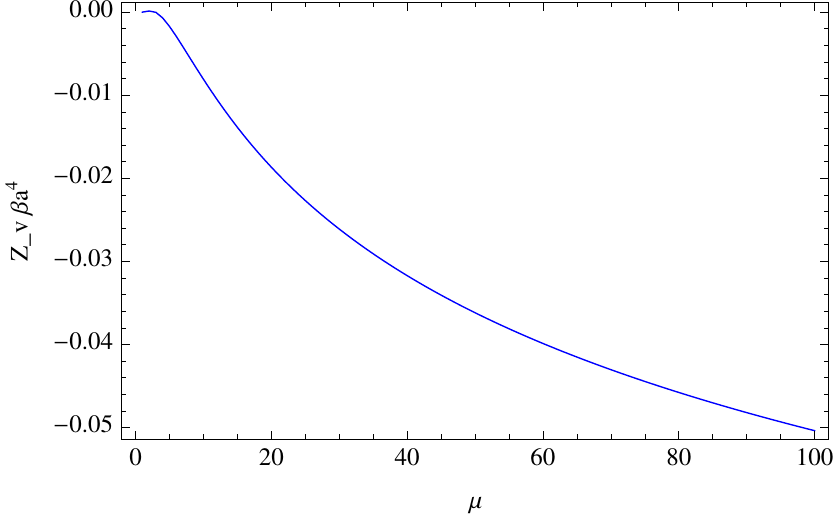}
\caption{Numerical evaluation of  $Z_v(a)$ (in units of $\beta a^4$) for the magnetic case as a function of $\mu$. $Z_v = 0$ for $\mu = 1$ and  diverges when $\mu \gg 1$, as 
expected for a 2+1 Neumann mode.} 
\label{fig2}
\end{figure}

In Fig.\ref{fig2}, we plot $Z_v$ in the case of magnetic materials as a
function of the permeability $\mu$. As described in Section \ref{magnetic},
$Z_v$ 
formally coincides  with $Z_s$ after the replacement $\epsilon\to 1/\mu$, so the numerical evaluation is similar to the previous one. $Z_v$ vanishes as $\mu \rightarrow 1$ and 
diverges for $\mu \gg 1$, as expected for a Neumann mode in 2+1 dimensions.

 
 Let us now consider the vector zero-mode contribution in   Eq.(\ref{f2vv}). The term proportional to
  $\sqrt{\Omega^2+\vert{\mathbf p}_\parallel+{\mathbf k}_\parallel\vert^2}$ does not generate inverse powers of $p_\parallel$ when expanded in powers of
  $k_\parallel$ (the plasma frequency acts as an infrared regulator). To analyze the last term, it is useful to consider the identity
  \begin{equation}
   \frac { ( {\mathbf p}_\parallel \cdot
({\mathbf p}_\parallel + {\mathbf k}_\parallel) )^2}{ |{\mathbf p}_\parallel | |{\mathbf
p}_\parallel + {\mathbf k}_\parallel| }={ |{\mathbf p}_\parallel | |{\mathbf
p}_\parallel + {\mathbf k}_\parallel| }-\frac{ |{\mathbf p}_\parallel| \sin^2\theta}{ |{\mathbf
p}_\parallel + {\mathbf k}_\parallel| } |{\mathbf k}_\parallel|^2\, .
\end{equation}
The second term in the above equation generates an infrared divergence
in the vector kernel. Indeed, when inserting this identity in Eq.(\ref{f2vv}), the term proportional to $\sin^2\theta$ reads, up to 
order $ |{\mathbf k}_\parallel|^2$,
\begin{equation}
\label{f2vvIR}
 -  \frac{2 |{\mathbf k}_\parallel|^2} {\beta}\, 
\int \frac{d^2{\mathbf p}_\parallel}{(2\pi)^2} \, 
\frac{\rho^2({\mathbf p}_\parallel)}{g^2({\mathbf p}_\parallel)}e^{-2\vert{\mathbf p}_\parallel\vert a }\sin^2\theta
\, .
\end{equation}
For small values of $p_\parallel$ we have $\rho^2\simeq 1$ and $g\simeq p_\parallel$, and therefore the integral above is logarithmically divergent
in the infrared.  This result shows that the NTLO correction to PFA for the zero-frequency vector mode is non-local as long as $\Omega$ is different
from zero. Unlike for magnetic materials, a finite value of $\Omega$ does not regulate the infrared divergence. This can be traced back to the fact
that the reflection coefficient $\rho^2\to 1$  as $p_\parallel\to 0$,  for any value of $\Omega$.

\section{Conclusions}\label{sec:conc}

In this paper we have analysed in detail the zero-frequency contribution to
the Casimir free energy in the presence of real materials.  We used a
functional approach, and clarified the use of the temporal gauge in the
context of the Casimir effect, since it that does not fix the zero frequency mode of $\widetilde A_0$.
This mode does in fact generate a TM contribution to the free energy.

We have also computed the kernels that are necessary to obtain the DE of
the Casimir free energy, in order to discuss the validity of the DE for
real materials. We have shown that the TM contribution does always produce a
NTLO correction to the PFA, which is local in derivatives of the function
$\psi$ that defines the shape of the curved interface. The same happens for
the TE contribution in the case of magnetic materials.  There is only one
situation where the DE would fail, that is, when
$\omega^2\epsilon(\omega)\to \Omega^2\neq 0$ as $\omega\to 0$ {\em for both
mirrors} \cite{footnote}.  
In terms of the models usually considered in the Casimir literature to describe real
materials, this condition  corresponds to the  plasma model. 

In summary, the non-analyticities we observed for perfect conductors in our
previous work \cite{deTfin}, survive only under the assumption of perfectly
lossless materials. Related with this, in recent works \cite{Guerout2014,
Guerout2015} it has been claimed that the Lifshitz-Matsubara formula does
not apply for the plasma model, and that it should be understood as the
lossless limit of the Drude model (note however that this claim has been
contested in Ref. \cite{Mostep2015CM}).  If this were the case, then
$\Omega^2$ would vanish, and there would be no  vector (or TE) contribution
to the zero mode. 
Consequently, the NTLO correction to PFA would be local even in this lossless limit.

\section*{Acknowledgements}
This work was supported by ANPCyT, CONICET, UBA and UNCuyo.
\appendix
\section{The scattering approach } \label{A}

In this Appendix we outline the calculation of $f^{(2)}=f_s^{(2)} + f_v^{(2)}$ using the scattering approach. In Ref.\cite{deothers} it is shown that
\begin{eqnarray}
\label{f2V}
&& f^{(2)}(k_\parallel,a) = -\frac{1}{\beta}\int \frac{d^2{\mathbf p}_\parallel}{(2\pi)^2}\sum_Q \frac{\vert{\mathbf p_\parallel}\vert r_Q(\pp)}{g_Q(\pp)}
e^{-2\vert\pp\vert a}\big[ (B_2)_{QQ}(\pp,\pp,\pp+\kp)
\nonumber\\
&+&2\sum_{Q'}\frac{\vert \pp+\kp\vert r_{Q'}(\pp+\kp) }{g_{Q'}(\pp+\kp)}e^{-2\vert\pp+\kp\vert a}
B_{QQ'}(\pp,\pp+\kp)B_{Q'Q}(\pp+\kp,\pp)\big]\, ,
\end{eqnarray}
where the indices $Q,Q'$ denote the two polarizations, $r_Q$ are the Fresnel reflection coefficients and
\begin{equation}
g_Q(\pp)=1-r_Q^2e^{-2\vert\pp\vert a}\, .
\end{equation}
We are assuming here that both media have the same EM properties.

The crucial ingredients in the above formula are the coefficients $B_{QQ'}$ and $(B_2)_{QQ}$, which describe the scattering 
of a classical EM wave at the rough interface between two homogeneous half-spaces. These coefficients have been
computed in Ref.\cite{Voronovich} using small-slope perturbation theory. The notation there is slightly different: 
$B_{QQ'}\to B_{\alpha \alpha_0}^{22}$ and
$(B_2)_{QQ}\to i (B_2)_{\alpha \alpha_0}^{22}$. The situation where both media have the same permittivity $\epsilon$ is obtained
by setting $\epsilon_1=\epsilon$ and $\epsilon_2=1$ in the results of Ref.\cite{Voronovich}.

As the general formula for $f^{(2)}$ has not been explicitly displayed in previous works, and is needed for the analysis of the emergence of non-analyticities
of the DE, we will work it out here. The coefficients  $B_{QQ'}$ are given in Eq.(4.15) of Ref.\cite{Voronovich}. One can check that, as we are considering
the zero-frequency mode, they  vanish for $Q\neq Q'$. Therefore, $f^{(2)}$ in Eq.(\ref{f2V}) has two separate contributions: $Q=Q'=1$ and $Q=Q'=2$, which 
correspond to the scalar $(f^{(2)}_s)$ and vector $(f^{(2)}_v)$ contributions, respectively. The coefficients $(B_2)_{QQ}$  are given in the Appendix D 
of Ref.\cite{Voronovich}.

\subsection{Scalar contribution}
The Fresnel reflection coefficient reads in this case 
\begin{equation}
r_1=\frac{\epsilon-1}{\epsilon +1}\, ,
\end{equation}
and the scattering coefficients
\begin{eqnarray}
\label{Bs}
B_{11}(\pp,\pp ') &=& -\frac{(\epsilon -1)}{(\epsilon +1)^2}\left[\epsilon +\frac{\pp\cdot\pp '}{\vert\pp\vert\vert\pp '\vert}\right]\\
(B_2)_{11}(\pp,\pp,\pp ') &=&  \frac{2(\epsilon -1)}{(\epsilon +1)^2\vert\pp\vert^2}\Big[ \frac{(\epsilon -1)}{(\epsilon+1)\vert\pp '\vert}
 (\epsilon \vert\pp\vert^2\vert\pp '\vert^2-(\pp\cdot\pp ')^2)+\frac{4\epsilon\vert\pp\vert}{(\epsilon+1)}\pp\cdot\pp '\Big]\nonumber .
\end{eqnarray}

Inserting Eq.(\ref{Bs}) into Eq.(\ref{f2V}) one obtains the scalar contribution to $f^{(2)}$. After a rather long calculation one can show that the result
coincides with Eq.(\ref{f2scal}).

\subsection{Vector contribution}
For $Q=2$, the Fresnel reflection coefficient reads
\begin{equation}
r_2=\frac{ |{\mathbf k}_\parallel| - \sqrt{|{\mathbf k}_\parallel|^2 + \Omega^2}}{|{\mathbf
k}_\parallel|+\sqrt{|{\mathbf
k}_\parallel|^2 + \Omega^2} } \;.
\end{equation}
Note that it coincides with the coefficient $-\rho$ defined in Eq.(\ref{rho}), in the particular case 
$\Omega_L=\Omega_R=\Omega$. 

On the other hand we have
\begin{eqnarray}
\label{Bv}
B_{22}(\pp,\pp ') &=& \frac{\Omega^2}{(\sqrt{\Omega^2+\vert\pp\vert^2}+\vert\pp\vert)(\sqrt{\Omega^2+\vert\pp '\vert^2}+\vert\pp '\vert)}
\frac{\pp\cdot\pp '}{\vert\pp\vert\vert\pp '\vert}
\nonumber\\
(B_2)_{22}(\pp,\pp,\pp ') &=& -2\rho(\pp)\left[ \frac{(\pp\cdot\pp ')^2}{\vert\pp\vert^2\vert\pp '\vert}+(\sqrt{\Omega^2+\vert\pp\vert^2}-\sqrt{\Omega^2+\vert\pp '\vert^2})\right]\, .
\end{eqnarray}

Inserting Eq.(\ref{Bv}) into Eq.(\ref{f2V}) one obtains the vector contribution to $f^{(2)}$, which is given in Eq.(\ref{f2vv}).


\begin{thebibliography}{bib}

\bibitem{books}P. W. Milonni, {\it The Quantum Vacuum}, Academic Press, San Diego, 1994; 
M. Bordag, G.L. Klimchitskaya, U. Mohideen, and V. M. Mostepanenko, {\it Advances in the Casimir Effect},
Oxford University Press, Oxford, 2009.

\bibitem{Mostep2015}G.~L.~Klimchitskaya and V.~M.~Mostepanenko,
  Proceedings of Peter the Great St.Petersburg Polytechnic
  University, N1 (517), pp.41-65, 2015.

\bibitem{Lifshitz}E. M. Lifshitz,  J. Exp. Theo. Phys. USSR{\bf 29}, 94 (1955), ibidem
Sov. Phys. JETP {\bf  2},  73 (1966).

\bibitem{Derjaguin}B.V. Derjaguin, Koll. Z. {\bf 69}, 155 (1934); B. V. Derjaguin and I. I. Abrikosova, Sov. Phys. JETP {\bf 3}, 819 (1957); 
B. V. Derjaguin, Sci. Am. {\bf 203}, 47 (1960).

\bibitem{denos}C.~D.~Fosco, F.~C.~Lombardo and F.~D.~Mazzitelli, Phys.\ Rev.\ D {\bf 84}, 105031 (2011).
See also  C.~D.~Fosco, F.~C.~Lombardo and F.~D.~Mazzitelli,
  Phys.\ Rev.\ A {\bf 89}, 062120 (2014),  and references therein.

\bibitem{deothers}G. Bimonte, T. Emig, R. Jaffe, and M. Kardar,  Europhys. Lett.  {\bf 97}, 50001 (2012);
 G. Bimonte, T. Emig, and M. Kardar, App. Phys. Lett. {\bf 100}, 074110 (2012).

\bibitem{Bimonte:2014mma} 
  G.~Bimonte, T.~Emig and M.~Kardar,
  Phys.\ Rev.\ D {\bf 90}, no. 8, 081702 (2014).

\bibitem{Lifshitzderiv}  I.E. Dzyaloshinskii, E.M. Lifshitz and L.P. Pitaevskii,
Sov. Phys. Uspekhi {\bf 4},  153 (1961);  C. Genet, A. Lambrecht and S. Reynaud, Phys. Rev. A {\bf 67},  043811 (2003); 
A. Lambrecht, P.A. Maia Neto and S. Reynaud, New J. Phys. {\bf 8},  243 (2006); 
C.~Ccapa Ttira, C.~D.~Fosco and F.~D.~Mazzitelli,  J.\ Phys.\ A {\bf 44}, 465403 (2011).  

\bibitem{Mostep2009}G.~L.~Klimchitskaya, U.~Mohideen and V.~M.~Mostepanenko,
  Rev.\ Mod.\ Phys.\  {\bf 81}, 1827 (2009).
  
  \bibitem{Guerout2014}R.~Gu\'erout, A.~Lambrecht, K.~A.~Milton and S.~Reynaud,
  Phys.\ Rev.\ E {\bf 90}, 042125 (2014).

\bibitem{deTfin}C.~D.~Fosco, F.~C.~Lombardo and F.~D.~Mazzitelli, Phys.\ Rev.\ D {\bf 86}, 045021 (2012).

\bibitem{Neumann2+1}C.~D.~Fosco, F.~C.~Lombardo and F.~D.~Mazzitelli,
  Phys.\ Rev.\ D {\bf 91}, 105019 (2015).

\bibitem{Voronovich} A.G. Voronovich, Waves in Random Media {\bf 4}, 337 (1994).

\bibitem{ZinnJustin} J.~Zinn-Justin, ``Quantum field theory and critical phenomena,'' Clarendon Press, Oxford (2002).

\bibitem{MiltonT}I.~H.~Brevik, S.~A.~Ellingsen and K.~A.~Milton, 
  New J.\ Phys.\  {\bf 8}, 236 (2006).
  
\bibitem{Ccappa} See C.~Ccapa Ttira et al, in Ref.\cite{Lifshitzderiv}.

\bibitem{footnote} If this happens for just one of the mirrors, there is no
contribution to the Casimir force between the mirrors, although the vacuum
energy itself is different.
  
\bibitem{Guerout2015}R.~Gu\'erout, A.~Lambrecht, K.~A.~Milton and S.~Reynaud, arXiv:1508.01659 [quant-ph].

\bibitem{Mostep2015CM} V.~M.~Mostepanenko,   Condens. Matter {\bf 27}, 214013 (2015).
  
\end{thebibliography}
\end{document}